\documentstyle[11pt]{article}
\newcommand{\be}{\begin{equation}}
\newcommand{\ee}{\end{equation}}
\newcommand{\ben}{\begin{eqnarray}}
\newcommand{\een}{\end{eqnarray}}

\newcommand{\no}{\noindent}
\newcommand{\n}{\label}
\begin{document}
\title{Inhomogeneous Universe Models with Varying \\
Cosmological Term}
\author{Luis P. Chimento$^{1}$\thanks{electronic address:
chimento@df.uba.ar}
$ \; $ and $ \; $ Diego Pav\'{o}n$^{2}$\thanks{electronic address:
diego@ulises.uab.es}\\
$^{1}$ Departamento de F\'{\i}sica, Facultad de Ciencias Exactas y Naturales\\
Universidad de Buenos Aires\\
Ciudad Universitaria, Pabell\'{o}n I, 1428 Buenos Aires, Argentina\\
$^{2}$ Departamento de F\'{\i}sica, Facultad de Ciencias\\
Universidad Aut\'onoma de Barcelona\\
08193 Bellaterra (Barcelona) Spain}
\date{}
\maketitle
\thispagestyle{empty}
\begin{abstract}
The evolution of a class of inhomogeneous spherically symmetric
universe models possessing a varying cosmological term and a material 
fluid, with an adiabatic index either constant or not, is studied. 
\end{abstract}
\newpage
One of the main puzzles concerning our current understanding of 
the physical world is the present small value of the 
effective cosmological
constant $\mid \Lambda \mid < 10^{-120}M_{Pl}^{2}$, 
or, which amounts to the same thing,
the small value of the vacuum energy density
$\rho_{v} = \Lambda/(8 \, \pi \, G)$, as witnessed by cosmic 
observation. The status of the problem was reviewed by Weinberg
\cite{WE}. A current of thought holds the view that the
cosmological term is not really constant but its value decreses 
as the universe expands. The rationale behind this is that the
energy of the vacuum should spontaneously decay into massive and
massless particles, hence reducing $\Lambda$ to a value compatible 
with astronomical constraints,
see e.g. \cite{OT}, \cite{FA}, \cite{GA}, \cite{ChW},
\cite{CO}, \cite{WA}, \cite{SW}, \cite{OW}, \cite{LM},
\cite{JM}, \cite{JN}, \cite{MP}. 
All these papers
adhere to one or another ``law" for the decay of $\rho_{v}$. However,
as shown by Pav\'{o}n \cite{DP} these laws 
should be  restricted by statistical 
physics considerations, and so several of them may be ruled out. 
All these studies were carried out on the assumption that 
the universe is homogeneous
and isotropic, which is to some extent very reasonable, 
for it is believed the universe has been so at least since 
shortly after the beginning of the expansion. To our knowledge 
the only work departing from a Robertson-Walker background 
is the one by Beesham who simultaneously considers $\Lambda$
and $G$ evolving with time in a Bianchi 
type-I universe \cite{BE}.

In this short article we consider a varying cosmological term in
a class of inhomogeneous (but isotropic) universe. The reason
to undertake such a study is that there is 
no reason {\it a priori} 
to believe that the universe was homogenoeous and isotropic
right back to the beginning of the expansion, and on the other 
hand there is the fact that
the homogeneous and isotropic models cannot account
for the high degree of homogeneity and isotropy so frequently
adscribed to the present state of our universe. Further 
motivations to study inhomegeneous cosmologies
can be found in \cite{KR} as well as in \cite{BEM}.

Let us consider the isotropic but inhomogeneous 
spherically-symmetric spacetime described  
by the plane Lema\^{\i}tre-Tolman-Bondi metric \cite{KR}
\be
ds^{2} = - dt^{2} + Y^{'2} \, dr^{2} + Y^{2}\, (d\theta^{2} + sin^{2}
\theta \, d\phi^{2}), \; \; (Y = Y(r, t)) 
\n{1}
\ee
(the ``prime" denotes partial derivative with respect to the radial 
coordinate $r$), 
the source of the metric being a material perfect fluid 
of energy density
$\rho(r, t)$ and pressure $P(r, t)$
plus the quantum vacuum.
The corresponding stress-energy tensor reads
\be
T_{ab} = (\rho + P ) u_{a} u_{b}
+ \left( P - \frac{\Lambda}{8 \, \pi \, G} \right) g_{ab} \, ,
\n{2}
\ee
where $u_{a} = \delta_{a}^{t}$ is the fluid fourvelocity. 
The nontrivial Einstein's equations take the form
\be
\rho + \Lambda = \frac{1}{Y^{2} \, Y^{'}} 
(\dot{Y}^{2} \, Y)^{'} \, ,
\n{3}
\ee 
\be
P - \Lambda = - \frac{1}{Y^{2} \dot{Y}} (\dot{Y}^{2}\, Y)^{.} \, ,
\n{4}
\ee
\be
\frac{\ddot{Y}}{Y} + \left( \frac{\dot{Y}}{Y}\right)^{2} -
\frac{\ddot{Y}^{'}}{Y^{'}}-\frac{\dot{Y}}{Y}\frac{\dot{Y}^{'}}{Y^{'}}=0 \, ,
\n{5}
\ee
where the upper dot means partial derivative with respect to time 
and we have set $ 8 \,  \pi \, G = 1 $. Introducing the change of
variables 
\[ Y = f^{2/3} \]
the last three equations become into
\begin{equation}
\rho + \Lambda = \frac{4}{3} \frac{ \dot{f} \, \dot{f}^{'}}
{f \, f^{'}} \, ,
\n{6}
\ee
\be
P - \Lambda = - \frac{4}{3} \frac{\ddot{f}}{f} \, ,
\n{7}
\ee
and
\be
\ddot{f}^{'} \, f - f^{'} \ddot{f} = 0,
\n{8}
\ee
respectively. 
We incorporate to this system the widely-used equation of state 
for the material fluid, namely
\be
P = (\gamma - 1) \rho
\n{9}
\ee
where very ofently the adiabatic index 
$\gamma$ is considered constant, 
though on physical grounds it may also depend on time. 
The latter possibility is very natural since if the 
quantum vacuum decays into 
a mixture of massive and massless particles, 
$\gamma$ must vary with time because both species
of particles redshift at different rates.
Equation (\ref{8}) readily implies  
\be
\ddot{f} - F(t) f = 0 \, ,
\n{10} 
\ee
where $F(t)$ does not depend on the radial coordinate. Note in passing
that, in the particular case, $\gamma = 1$ equations (\ref{7}) and
(\ref{10}) lead to $ \Lambda = \Lambda (t)$. For $\gamma \neq 1$
we will have in general that $\Lambda = \Lambda (r, t)$, 
though $\Lambda = \Lambda (t)$ is also possible.

Introducing the factorization
\[ f(r, t) = R(r) \, T(t) \]
in (\ref{10}) we get
\be
\ddot{T} - F \, T = 0.
\n{11}
\ee
The latter has two independent solutions, and therefore 
the general solution to (\ref{10}) is
\[
f(r, t) = R_{1} \, T_{1} (t) + R_{2} \, T_{2} (t) \, .
\] 
However we will restrict ourselves to the simpler case
\be 
f(r, t) = R(r) \, T(t) \, ,
\n{12}
\ee 
which is nonetheless fairly general. In this way (\ref{8})
becomes an identity and equations (\ref{6}) and (\ref{7})
reduce to
\be 
\rho = \frac{4}{3} \, \left(\frac{\dot{T}}{T} \right)^{2} 
- \Lambda
\n{13}
\ee 
and
\be 
P = - \frac{4}{3} \, \frac{\ddot{T}}{T} + \Lambda \, ,
\n{14} 
\ee 
respectively. Note that because of the above factorization any
regular function $R(r)$ will satisfy (\ref{10}).  
To get an equation for $T(t)$ we  
substitute the right hand sides of (\ref{13}) and (\ref{14}) 
in (\ref{9}); it follows
\be
T \ddot{T} + (\gamma - 1) \,  \dot{T}^{2} - \frac{3}{4}
\Lambda \, T^{2} = 0 \, . 
\n{15}
\ee
To solve it we resort to the change of variables
\be 
T = Z^{n}  \; \; \; \; \; ( n = \mbox{constant}) \, ,
\n{16}
\ee
which leads to $ n = 1/\gamma$ and 
\be 
\ddot{Z} - \frac{3}{4} \gamma \Lambda \, Z = 0 \, .
\n{17}
\ee 
Depending on the expression of 
$\Lambda$, which in what follows will be assumed 
position independent (the latter automatically implies that
both $\rho$ and $P$ will depend on $t$ only),
different cases arise.

\begin{enumerate} 
\item For $\gamma$ and $\Lambda$ constants one obtains
\ben
Z_{1} = C_{1} \, cosh \, \left( \frac{\sqrt{3 \gamma \Lambda}}{2} \; 
t + \varphi_{1} \right) \, , \\
\n{18}
Z_{2} = C_{2} \, sinh \, \left( \frac{\sqrt{3 \gamma \Lambda}}{2} \;
t + \varphi_{2} \right) \, ,
\n{19}
\een 
and therefore
\be
Y_{1} = R^{2/3} (r) \, C_{1}^{2/3\gamma} \, cosh^{2/3\gamma} \left(
\frac{\sqrt{3 \gamma \Lambda}}{2} \; t + \varphi_{1} \right) \, , \\
\n{20}
\ee 
\be
Y_{2} = R^{2/3} (r) \, C_{2}^{2/3\gamma} \, sinh^{2/3\gamma} \left(
\frac{\sqrt{3 \gamma \Lambda}}{2} \; t + \varphi_{2} \right).
\n{21}
\ee 
Note that solution (\ref{20}) does not present initial singularity,
but solution (\ref{21}) has a singularity at
$ t_{0} = - \frac{ 2 \varphi_{2}}{\sqrt{ 3 \gamma \Lambda}}$.
However both sets of solutions have a final inflationary stage.

\item For $\gamma = \mbox{constant}$ and 
\be
\Lambda (t) = \frac{\lambda_{0}^{2}}{t^{2}} \;  \; \; 
(\lambda_{0}^{2} = \mbox{constant}) \, ,
\n{22}
\ee 
the corresponding differential equation can be integrated by
using the {\em ansatz} $Z \propto t^{m}$ with $m =$ constant.
The general solution reads
\be 
Z(t) = C_{1} \, t^{m_{+}} + C_{2} \, t^{m_{-}} 
\n{23}
\ee 
where $ m_{\pm} = \frac{1}{2} \,\pm \, \frac{\sqrt{1 + 3 \gamma
\lambda_{0}^{2}}}{2}$. Inflationary solutions may occur for large 
enough $\lambda_{0}^{2}$.

\item For $\gamma = \mbox{constant}$ and 
\be
\Lambda = \lambda_{0}^{2} \, t^{n-2}  \; \; \; (n \neq 0, \, 2) \, ,
\label{24}
\ee
equation (\ref{17}) becomes
\be 
\ddot{Z} - \frac{3}{4} \gamma \, \lambda_{0}^{2} \, 
t^{n-2}\, Z = 0
\n{25}
\ee 
and the general solution can be expressed as a combination of Bessel 
functions (see Ref. 18)

$$
Z = C_{1} \, t^{1/2} \, J_{1/n}\left(\frac{\lambda_{0}}{n}
\sqrt{- 3\gamma} \, t^{n/2} \right) 
$$

\be 
+ \, C_{2} \, t^{1/2} \, J_{-1/n}\left(\frac{\lambda_{0}}{n}
\sqrt{-3\gamma} \, t^{n/2} \right).
\n{26}
\ee 
The behavior at the asymptotic limits depends on $n$. 
For $0 < n <2$ one has the following.
 (i) When $ t \to 0$ one obtains

\be
\n{27}
Z \sim C_{1} \, t + C_{2} \, .
\ee

\no One can choose $C_{2} = 0$ to
have the initial singularity at $t = 0$. (ii) When $t \to \infty$
there follows

\be
\n{28}
Z \sim t^{\frac{1}{2}-\frac{n}{4}} \; cos \, t^{n/2}.
\ee

\no For $ n < 0 $ one has the following: (i) when $ t \to 0 \; $
one obtains
$Z \sim t^{\frac{1}{2}-\frac{n}{4}} \; cos \, (t^{n/2} + \varphi)\;. $
(ii) When $ t \to \infty \; $ one obtains
$ Z \sim t \, .$

\item For $\gamma = \mbox{constant}$ and

\be
\Lambda (t) =\lambda_{0}^{2}+c\mbox{e}^{-\alpha t}
\n{29}
\ee

\no where $\lambda_{0}^{2}$, $c$ and $\alpha$ are constants
(with $c < 0 $ for mathematical convenience), equation
(\ref{17}) becomes

\be 
\ddot{Z} - \frac{3}{4} \gamma \,\left[\lambda_{0}^{2}+c\mbox{e}^{-\alpha
t}\right]Z = 0
\n{30}
\ee 

and the general solution can be expressed as a combination of Bessel 
functions (see Ref. 18)

$$
Z = C_{1} \,J_{\frac{\lambda_{0}}{\alpha}\sqrt{3\gamma}}
\left(\frac{\sqrt{- 3\gamma c}}{\alpha}\,\mbox{e}^{\frac{-\alpha}{2} t}\right)
$$

\be 
+ \, C_{2} \,J_{-\frac{\lambda_{0}}{\alpha}\sqrt{3\gamma}}
\left(\frac{\sqrt{- 3\gamma c}}
{\alpha}\,\mbox{e}^{\frac{-\alpha}{2} t}\right) \, ,
\n{31}
\ee 
with

\be
\n{32}
C_2=-\frac{J_{\frac{\lambda_{0}}{\alpha}\sqrt{- 3\gamma}}
\left(\frac{\sqrt{- 3\gamma c}}{\alpha}\right)}
{J_{-\frac{\lambda_{0}}{\alpha}\sqrt{- 3\gamma}}
\left(\frac{\sqrt{- 3\gamma c}}{\alpha}\right)}\,C_1
\ee

\no in order to fix the initial singularity at $t=0$.

\no The asymptotic behavior near the initial singularity, when $t\to 0$,
is given by

\be
\n{33}
 Z \sim \, t.
\ee

\no  At the final stage, when $t\to\infty$ 
and $\Lambda\to\lambda_0^2$, one
obtains the following asymptotic behavior

\be
\n{34}
Y\approx R^{2/3}(r)\,\mbox{e}^{\frac{\lambda_{0}}
{\sqrt{-3\gamma c}}\,t} \, .
\ee

\no Besides, from (\ref{21}) we recover the same 
result in the far future.

\no For the particular case $\lambda_0^2=0$ the general 
solution of (\ref{30}) is given by

\be
Z = C_{1}\,J_0\left(\frac{\sqrt{- 3\gamma c}}{\alpha}\,
\mbox{e}^{\frac{-\alpha}{2} t}\right)\, +\,
C_{2}\,Y_0\left(\frac{\sqrt{- 3\gamma c}}{\alpha}\,
\mbox{e}^{\frac{-\alpha}{2} t}\right) \, ,
\n{35}
\ee 

\no where $Y_{0}$ is the Weber function of the 
second kind and zero order. In
the limit $t\to\infty$ and $\Lambda\to 0$ the final 
behaviour of the solutions, obtained from (\ref{33}) are

\be
\n{36}
Y\approx R^{2/3}(r)\,t^{2/3\gamma} \, .
\ee

\no The same result can be easily obtained from (\ref{17}) 
by setting $\Lambda=0$.

\item For $\gamma = \gamma (t)$ and $\Lambda = \Lambda (t)$
we shall find the expressions of both quantities 
and analyse the behavior of the solutions at late time. To do
this we introduce the new varible $s(t)$ in the following way
\be
\n{37}
T=T_{0}\,\mbox{e}^{\int{\frac{1}{\gamma}\frac{\dot s}{s}\,dt}},
\ee
\no inserting (\ref{37}) into (\ref{15}) we get
\be
\n{38}
\ddot s-\frac{\dot\gamma}{\gamma}\dot s+\frac{3}{4}\Lambda\gamma s=0.
\ee
\no This equation can be identified with
\be
\n{39}
\ddot{s}+s^{n}\dot{s}+\frac{1}{(n+2)^{2}}s^{2n+1}=0  \qquad   (n\ne -2),
\ee
\no which is reduced to a linear differential equation by 
making the substitution
\cite{luis}
\begin{equation}
\n{40}
s^{n} = \frac{n+2}{n} \frac{v^{n}}{c_{1} + \int{v^{n}\,dt}},
\end{equation}
\noindent obtaining
\begin{equation}
\n{41}
\ddot{v}=0, \qquad  v(\tau)=c_{2}+c_{3}\,t,
\end{equation}
\no where $c_1$, $c_2$ and $c_3$ are arbitrary integration constants.
Equations (\ref{38}) and (\ref{39}) are the same if we define
\be
\n{42}
-\frac{\dot\gamma}{\gamma}=s^n,
\ee
and
\be
\n{43}
\frac{3}{4}\Lambda\gamma=\frac{1}{(n+2)^{2}}s^{2n}.
\ee
\no Without loss of generality we choose $c_2=-t_0$, where $t_{0}$ 
is some initial time, and $c_3=1$. So, the last
system of equations can be easily solved to obtain
\be
\n{44}
\Lambda(t)=\frac{4C^2(t-t_0)^{2n}}{3\gamma_0 n^2(n+1)^2}
\left[1+\frac{(t-t_0)^{n+1}}{C}\right]^{\frac{2-n}{n}},
\ee
and
\be
\n{45}
\gamma(t)=\gamma_0\left[1+\frac{(t-t_0)^{n+1}}{C}\right]^{-\frac{2+n}{n}},
\ee
\no where $\gamma_0$ and $C$ are arbitrary integration constants. 
On the other hand, inserting (\ref{41}) in (\ref{40}), 
the general solution of the nonlinear equation (\ref{39}) is
found to be
\be
s(t)=\left[\frac{(n+1)(n+2)}{n}\frac{(t-t_0)^n}{C+(t-t_0)^{n+1}}
\right]^{1/n} \,.
\n{46}
\ee
\no Now, taking into account that at late time, $t\gg t_0$, 
we must have $\gamma\to\gamma_0$, the restriction $n<-1$ 
readily follows as can be seen from (\ref{45}). In addition the
cosmological term vanishes in the same limit. Now, using 
this approximation we evaluate $T(t)$ in (\ref{37}), finding
\be
\n{47}
T(t)\approx T_0\left[\frac{(n+1)(n+2)}{n}(t-t_0)\right]^{1/\gamma_0}
\, ,
\ee
and therefore
\be
\n{48}
Y\approx R^{2/3} (r) \, T_{0}^{2/3\gamma_0} \,
\left[\frac{(n+1)(n+2)}{n}(t-t_0)\right]^{2/3\gamma_0} \, .
\ee
\no It is worthy of note that, for $t\gg t_0$ we
have both $\gamma\to\gamma_0$ and $\Lambda\to 0$. So, using these
limits in (\ref{21}) we recover the solution given by (\ref{48}).

\end{enumerate}

To investigate the singular structure of the plane 
Lema\^{\i}tre-Tolman-Bondi metric (\ref{1}), we calculate 
the curvature scalar by resorting to change of 
variables $Y=f^{2/3}$ used above
\be
\n{49}
{\cal R} =2\,\frac{\ddot f}{f}+\frac{4}{3}\,
\frac{\dot f\,\dot f^{'}}{f\,f^{'}} + 2 \frac{\ddot f^{'}}{f'} \, ,
\ee

\no and evaluate it at the points where the 
coefficients of the metric $Y'^2$
and/or $Y^2$ vanish. To do this we insert the Einstein 
equation (\ref{8})
along with (\ref{12}), (\ref{15}) and 
(\ref{16}) in (\ref{49}), obtaining
\be
\n{50}
{\cal R} =4\,\left[\frac{4}{3}\Lambda+\frac{(2-\gamma)}{\gamma^2}
\frac{\dot Z^2}{Z^2}\right].
\ee

\no Then we replace in the scalar curvature the expansion 
of $\Lambda$ and the corresponding solutions near the 
point where they vanish. All the solutions we
have found for $\gamma=$ constant, except (\ref{20}), 
have a singularity at $t = 0$, i.e. the big-bang singularity.

In summary we have found the coefficients of the 
Lema\^{\i}tre-Tolman-Bondi metric assuming that the early
universe possessed a time varying cosmological term, and
that the adiabatic index of the material fluid were either 
constant or not. (i) All the solutions we have derived contain
an arbitrary function of the radial coordinate. (ii) For 
$\gamma = $ constant all the solutions, except (\ref{20})
have a singularity at $t = 0$, i.e. the big-bang singularity.
(iii) Constant as well as varying cosmological terms give
rise asymptotically to exponential inflation -see (\ref{20}),
(\ref{21}) and (\ref{34}). (iv) For a varying cosmological
term there exist solutions which behave as though 
the universe were asymptotically matter dominated 
at late times when $\gamma = 1$ -see (\ref{36}). 

None of the solutions found has a spatially-homogeneous limit for 
$t \to \infty$. This is so because no homogeneization mechanism,
such as anisotropic pressures \cite{RS}, was assumed. Such 
a more general study will be undertaken soon. 

This work was partially supported by the Spanish Ministry 
of Educationunder Grant PB94-0718, and the University of 
Buenos Aires under Grant EX-260.

\end{document}